# Algebraic Extension Ring Framework for Non-Commutative Asymmetric Cryptography

P. Hecht[1]

*Abstract*— Post-Quantum Cryptography (PQC) attempts to find cryptographic protocols resistant to attacks using Shor's polynomial time algorithm for numerical field problems or Grover's algorithm to find the unique input to a black-box function that produces a particular output value. The use of non-standard algebraic structures like non-commutative or non-associative structures, combined with one-way trapdoor functions derived from combinatorial group theory, are mainly unexplored choices for these new kinds of protocols and overlooked in current PQC solutions. In this paper, we develop an algebraic extension ring framework who could be applied to different asymmetric protocols (i.e. key exchange, key transport, enciphering, digital signature, zero-knowledge authentication, oblivious transfer, secret sharing etc.). A valuable feature is that there is no need for big number libraries as all arithmetic is performed in $\mathbb{F}_{256}$ extension field operations (precisely the AES field). We assume that the new framework is cryptographical secure against strong classical attacks like the sometimes-useful length-based attack, Roman'kov's linearization attacks and Tsaban's algebraic span attack. This statement is based on the non-linear structure of the selected platform which proved to be useful protecting the AES protocol. Otherwise, it could resist post-quantum attacks (Grover, Shor) and be particularly useful for computational platforms with limited capabilities like USB cryptographic keys or smartcards. Semantic security (IND-CCA2) could also be inferred for this new platform.

## 1. Introduction

Post-Quantum Cryptography (PQC) is a trend that has an official NIST status [1] and which aims to be resistant to quantum computers attacks like Shor [2] and Grover [3] algorithms. NIST initiated a process to solicit, evaluate, and standardize one or more quantum-resistant public-key cryptographic algorithms. Particularly Shor algorithm provided a quantum computing way to break asymmetric protocols.

PQC not only cover against that menace, it works also as a response against side-channel attacks [3], the increasing concern about pseudo-prime generator backdoor attacks (i.e. Dual_EC_DRBG NSA [4]) or the development of quasi-polynomial discrete logarithm attacks which impact severely against current de facto standards [5, 6, 7] of asymmetric cryptography whose security rest on integer-factorization (IFP) and discrete-logarithm (DLP) over numeric fields.

As a response, there is a growing interest in PQC solutions like Lattice-based, Pairing-based, Multivariate Quadratic, Code-based and Hash-based cryptography [8, 9, 10], Another kind, and overlooked solutions belong to Non-Commutative (NCC) and Non-Associative (NAC) algebraic cryptography [11-15].

Security of a canonical algebraic asymmetric protocol always relies on a one-way function (OWF) transformed to work as a one-way trapdoor function (OWTF) [11, 16]. For instance, using the decomposition problem (DP) or the double coset problem (DCP) [8, 9, 10], both assumed to belong to AWPP time-complexity (but out of BQP) [17, 18] problems, which lead to an eventual brute-force attack, thus yielding high computational security. A solution which does not require commutative subgroups is the Anshel-Anshel-Goldberg (AAG) key-exchange protocol (KEP) [11].

PQC studies were purposely followed by the author over his past and current research [19-23].

## 2. The Algebraic Extension Ring framework

The Algebraic Extension Ring (AER) framework include the following structures:

$\mathbb{F}_{256}$: AES field [24, 25]
    Primitive polynomial: $1+x+x^3+x^4+x^8$
    Multiplicative subgroup generator: <1+x>

$M[\mathbb{F}_{256}, n]$ n-dimensional square matrix of field elements. (bytes). Each matrix is equivalent to a (rank-3) Boolean tensor. Matrix powers are obtained through usual rules with field sum and field product operations. Elements do not commute. Table I show the alternative views of a random element.

The AER platform has two substructures:

$(M[\mathbb{F}_{256}, n], \oplus, O)$    Abelian group using field sum as operation and null matrix as the identity element.

$(M[\mathbb{F}_{256}, n], \odot, I)$    Non-commutative monoid using field product as operation and identity matrix as the identity element.

The product distributes over sum of elements. Among AER elements, some are cyclic multiplicative subgroup generators, the rest have periodic power sets that do not include identity matrix (see Table II). This second class has therefore non-identity spurious inverses. Among matrices, random exploration search revealed that the proportion of cyclic subgroup generators decrease with increasing dimension. The multiplicative order of that kind of matrices and the conditions under which the cyclic subgroup feature appears, are open questions that should be further studied.

The inverses of each element are obtained by computing the multiplicative order of each one, no general algorithm is known, and this fact could represent a cryptographic advantage because obtaining inverses of AER elements

---

[1] Pedro Hecht: Maestría en Seguridad Informática, FCE-FCEyN-FI (UBA-Argentina) qubit101@gmail.com

require an unrestrained exploration.

TABLE I
REPRESENTATION OF A 2-DIMENSIONAL AER-RING ELEMENTS.

Decimal view = $\begin{pmatrix} 183 & 62 \\ 77 & 50 \end{pmatrix}$,

Rank – 3 Tensor view =
$\begin{pmatrix} \{1, 1, 1, 0, 1, 1, 0, 1\}_2 & \{0, 1, 1, 1, 1, 1, 0, 0\}_2 \\ \{1, 0, 1, 1, 0, 0, 1, 0\}_2 & \{0, 1, 0, 0, 1, 1, 0, 0\}_2 \end{pmatrix}$,

Polynomial view =
$\begin{pmatrix} 1 + x + x^2 + x^4 + x^5 + x^7 & x + x^2 + x^3 + x^4 + x^5 \\ 1 + x^2 + x^3 + x^6 & x + x^4 + x^5 \end{pmatrix}$,

TABLE II
POWER SET OF AN AER-ELEMENT WITH SPURIOUS IDENTITY

$\{\begin{pmatrix} 165 & 182 \\ 199 & 138 \end{pmatrix}, \begin{pmatrix} 110 & 217 \\ 146 & 87 \end{pmatrix}, \begin{pmatrix} 35 & 213 \\ 242 & 62 \end{pmatrix},$
$\begin{pmatrix} 230 & 10 \\ 80 & 208 \end{pmatrix}, \begin{pmatrix} 42 & 61 \\ 243 & 153 \end{pmatrix}, \begin{pmatrix} 170 & 161 \\ 127 & 224 \end{pmatrix},$
$\begin{pmatrix} 192 & 210 \\ 202 & 200 \end{pmatrix}, \begin{pmatrix} 95 & 199 \\ 98 & 60 \end{pmatrix}, \begin{pmatrix} 93 & 146 \\ 252 & 142 \end{pmatrix},$
$\begin{pmatrix} 3 & 242 \\ 209 & 235 \end{pmatrix}, \begin{pmatrix} 113 & 80 \\ 182 & 218 \end{pmatrix}, \begin{pmatrix} 75 & 243 \\ 217 & 164 \end{pmatrix},$
$\begin{pmatrix} 39 & 127 \\ 213 & 65 \end{pmatrix}, \begin{pmatrix} 90 & 202 \\ 10 & 26 \end{pmatrix}, \begin{pmatrix} 206 & 98 \\ 61 & 251 \end{pmatrix},$
$\begin{pmatrix} 222 & 252 \\ 161 & 28 \end{pmatrix}, \begin{pmatrix} 24 & 209 \\ 210 & 25 \end{pmatrix}, \begin{pmatrix} 165 & 182 \\ 199 & 138 \end{pmatrix}\}$

Table II: successive power list (1 to 18) of a random element (orange red) that generate a 17-period ending in a spurious identity (blue magenta) and the previous one (yellow) is a suprious inverse.. This is the reason why the AER structure is a ring but not a field, since some elements have no multiplicative order.

### 3. CRYPTOGRAPHIC ASPECTS

The choice of the $\mathbb{F}_{256}$ field (AES) [24] to generate the AER-structure is a central feature of this framework because it assures strong confidence that no linearization attack against the AER framework could succeed, i.e. Roman'kov three described types [26] or Tsaban Algebraic Span Attack [27].

Security of a canonical asymmetric cipher protocol always relies on a one-way Trapdoor Function (OWTF) [16]. Here we propose Public-Key Cryptography (PKC) protocols selecting algebraic schemes as the one-way trapdoor function. If the algebraic structure and OWTF are well selected, a provably secure protocol could be developed [11]. This sounds simple, but it is not easy to prove such a claim [28, 29]; so, caution at use is strongly advised. Commonly used OWTF in non-commutative cryptography (NCC) are, among others [11]:

1. (CSP) The conjugacy search problem given a recursive presentation of a group G and two elements g, h ∈ G, find out whether there is an element x ∈ G, such that $x^{-1}g\,x = h$.

2. (DSP) The decomposition search problem given a recursive presentation of a group G, two recursively generated subgroups A, B ⊆ G, and two elements g, h ∈ G, find two elements x ∈ A and y ∈ B that would satisfy x g y = h, provided at least one such pair of elements exists.
3. (DCP) The double coset problem is a special case of the decomposition search problem, where A=B.
4. (XTDP) The extended triple decomposition problem: a strong extension of Kurt's Triple Decomposition Problem who up to date resists the most advanced algebraic span linearization attack [27]. XTDP is described and applied elsewhere [23].

In this paper, we use the Anshel-Anshel-Goldberg (AAG-KEP) protocol for a key-exchange protocol [11]. The advantage of AAG is that it does not require commutative subgroups of a non-commutative platform, it requires only efficiently solvable word problem of any non-abelian algebraic structure. This implies an efficient inverse computation, an open question for the present framework which has a singular derived solution for 2-dimensional AER.

### 4. AAG-KEP

Let G be a public non-commutative structure that has an efficient resolution for the WP (word problem) and public elements A: $\{a_1,…, a_k\}$ and B: $\{b_1,…, b_m\}$. The protocol security relies on two problems: SCSP (simultaneous conjugation search problem) and MSP (membership search problem). Conjugation is represented as a power: $x^y = y^{-1}x\,y$

(a) Alice chooses (at random) a private element x ∈ G as a word in A, that is $x = f(a_1, ..., a_k)$ and sends ($b_1^x$, ..., $b_m^x$) to Bob.
(b) Bob chooses (at random) a private element y ∈ G as a word in B, that is $y = g(b_1,…, b_m)$ and sends ($a_1^y$,…, $a_k^y$) to Alice.
(c) Alice computes $f(a_1^y,…, a_k^y) \equiv x^y = y^{-1}xy$ and pre-multiplies it by $x^{-1}$.
(d) Bob computes $g(b_1^x,…, b_m^x) \equiv y^x = x^{-1}yx$, pre-multiplies it by $y^{-1}$, finally invert the obtained word leaving $x^{-1}y^{-1}xy$
(e) Both end with the same key, the commutator [x, y].

### 5. COMPUTING INVERSE ELEMENTS IN 2-DIMENSION AER

No polynomial time algorithm is known for computing directly the inverses of AER elements, but a sequential search could work in case of low dimensions. The strategy is based on the forced periodic behavior of all power sets. In AER, we empirically found that the first repeating element, so periods are not preceded by an aperiodic sequence. Therefore, if we find the period (p) of a power set of a base (x) AER element who ends with the identity (x, $x^2$, $x^3$,…, $x^p$=1), the penultimate of the set is an inverse of the considered element ($x^{p-1}=x^{-1}$). In numeric fields, any power set end always with the multiplicative identity but in AER this is not true and spurious identities appears (Table II), in this case no multiplicative order could be derived and his previous element of the power set sequence act as an spurious inverse.

To obtain the period (p) of the power set, we could use Floyd algorithm [30] or Floyd-Brent algorithm [31]. Computing powers could be efficiently solved using the fast square-and-multiply algorithm [16].

For 2-dimensional AER, a non-sequential shortcut allows fast inverse computation without previous period determination. As powers of any order could be efficiently computed and pigeonhole principle predicts that the cardinal of 2-dimensional AER-elements (=$256^4$-1) is unavoidably a multiple of any period and the $256^4$-2 power of any element (x) is an inverse ($x^{-1}$). This solves the low dimensional cases but for higher dimension this approach becomes impractical, as Table III shows.

TABLE III
CARDINALITY OF AER SETS

| Dimension | Cardinal |
| --- | --- |
| 2 | $4.294967296 \times 10^9$ |
| 3 | $4.722366482869645 \times 10^{21}$ |
| 4 | $3.402823669209384 \times 10^{38}$ |
| 5 | $1.60693804425899 \times 10^{60}$ |
| 6 | $4.973232364097866 \times 10^{86}$ |
| 7 | $1.008691358627698 \times 10^{118}$ |
| 8 | $1.340780792994259 \times 10^{154}$ |

We define a tensor determinant as the generalized determinant of an AES element using field sum and product instead of arithmetic operations.

Recent results show that spurious inverses and identities appears when the tensor determinant of the tested AES element is null. We call this case a singular tensor.

A step forward to develop an efficient computation of the periods of AES elements is to use the divisors list of the respective cardinality minus one (see Table III). Each period of a cyclic subgroup (and spurious behavior) must be a divisor of the multiplicative group size (Lagrange), trials of increasing divisors are used until the first identity (or spurious identity) appears.

## 6. STEP-BY-STEP EXAMPLE

The Appendix present an example of the AAG-KEP worked out using AER elements and Mathematica language. Black text corresponds to language input, blue text the corresponding output, in red there are comments and in magenta the obtained shared key. The new derived functions are: rmat - a random AER element, TProd[x,y] - field product of x and y AER elements and TFastPower[x,n] - the power $x^n$. Inverses of elements are obtained in one-step using a power function with the defined limit as the special exponent explained in the previous section, a forced multiple of the multiplicative order.

## 7. PROTOCOL SECURITY

AAG is clearly a PQC protocol, resisting currently defined quantum algorithms. Classical attacks, i.e. the length-based attack, are described elsewhere [Myasnikov]. If no polynomial time attack algorithm work, security relies on the dimension used (see Table III.). For real life application it is suggested to use A and B sets with at least 100 elements each. Overall security could be increased if an efficient inverse computation algorithm could be derived for elements of higher dimension, but it is also possible to use 2 or 3-dimensional AER elements if the public sets (A, B) are of considerable size, making unfeasible classical cryptanalytic attacks.

## 8. SEMANTIC SECURITY

Random elements of the AER structure are achieved using uniformly random bytes. It is obvious that random elements have no bias and therefore are all indistinguishable. Field operations (sum product, powers and inverses) do not generate special subsets of elements that could segregate shared keys from any other random element. Therefore, it could be inferred that this protocol has semantic security at IND-CCA2 level [32, 33].

## 9. CONCLUSIONS

At present work, a new kind of non-commutative framework is presented as a secure platform for canonical asymmetric protocols. The main advantage is the use a matricial ring with AES field operations instead of ordinary arithmetic, providing assurance against linearization and algebraic attacks. A full 2-dimensional elements implementation is presented and if polynomial time resolution of inverses computation could be derived, this could be generalized to higher dimensions. Some arguments are given to support the semantic security of the framework.

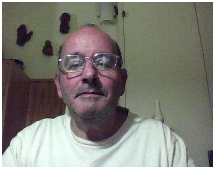
**Pedro Hecht** received an MSci in Information Technology at Escuela Superior de Investigación Operativa (ESIO-DIGID) and an PhD degree from Buenos Aires University (UBA). Currently, he is full professor of cryptography at Information Security Graduate School at UBA, EST (Army Engineering School), IUPFA (Federal Police University) and ENI (National Intelligence School). He works also as an UBACyT research fellow and Director of EUDEBA (UBA editorial board).

# APPENDIX: AAG-KEP EXAMPLE USING AER ELEMENTS

```
step 0 (SETUP)
dim = 2;
limit = (256^(dim^2)) - 2;

(* DEFINE A,B random sets of 5 AER elements each *)
A = {rmat, rmat, rmat, rmat, rmat};
B = {rmat, rmat, rmat, rmat, rmat};

Table[A[[i]] , {i, 1, 5}]
{{{234, 67}, {219, 0}}, {{162, 202}, {121, 143}},
 {{67, 137}, {220, 106}}, {{199, 110}, {183, 92}},
 {{237, 239}, {211, 252}}}
Table[B[[i]] , {i, 1, 5}]
{{{88, 183}, {153, 25}}, {{8, 73}, {160, 26}},
 {{142, 10}, {22, 153}}, {{202, 231}, {110, 0}},
 {{47, 118}, {238, 49}}}

step 1
(* DEFINE x =f(A)=a1 . (a3)^2 . (a5)^-1 *)
{x1 = A[[1]],
 x2 = TProd[A[[3]], A[[3]]],
 x3 = TFastPower[A[[5]], limit]}
{{{234, 67}, {219, 0}}, {{157, 61}, {184, 176}},
 {{139, 111}, {158, 137}}}
x = TProd[TProd[x1, x2], x3]
{{244, 199}, {161, 106}}
invx = TFastPower[x, limit]
{{207, 42}, {35, 163}}
APrime = Table[TProd[TProd[invx, B[[i]]], x], {i, 1, 5}]
{{{96, 46}, {247, 33}}, {{201, 184}, {199, 219}},
 {{181, 239}, {202, 162}}, {{157, 134}, {43, 87}},
 {{57, 198}, {39, 39}}}

step 2
(* DEFINE y=g(B)=(b1)^3 . (b2)^-2 . b4 *)
{y1 = TProd[TProd[B[[1]], B[[1]]], B[[1]]],
 y2 = TProd[TFastPower[B[[2]], limit], TFastPower[B[[2]], limit]],
 y3 = B[[4]]}
{{{105, 152}, {218, 62}}, {{96, 185}, {146, 230}},
 {{202, 231}, {110, 0}}}
y = TProd[TProd[y1, y2], y3]
{{54, 252}, {233, 201}}
invy = TFastPower[y, limit]
{{167, 247}, {222, 209}}
BPrime = Table[TProd[TProd[invy, A[[i]]], y], {i, 1, 5}]
{{{155, 88}, {93, 113}}, {{67, 219}, {82, 110}},
 {{70, 38}, {195, 111}}, {{45, 184}, {255, 182}},
 {{199, 175}, {205, 214}}}

step 3
{xprime1 = BPrime[[1]],
 xprime2 = TProd[BPrime[[3]], BPrime[[3]]],
 xprime3 = TFastPower[BPrime[[5]], limit]}
{{{155, 88}, {93, 113}}, {{16, 161}, {146, 61}},
 {{58, 176}, {107, 56}}}
xprime = TProd[TProd[xprime1, xprime2], xprime3]
{{138, 127}, {241, 20}}
KEYalice = TProd[invx, xprime]
{{136, 128}, {80, 156}}

step 4
{yprime1 = TProd[TProd[APrime[[1]], APrime[[1]]], APrime[[1]]],
 yprime2 = TProd[TFastPower[APrime[[2]], limit],
 TFastPower[APrime[[2]], limit]],
 yprime3 = APrime[[4]]}
{{{64, 66}, {133, 23}}, {{226, 245}, {221, 100}},
 {{157, 134}, {43, 87}}}
yprime = TProd[TProd[yprime1, yprime2], yprime3]
{{70, 1}, {182, 185}}
invKey = TProd[invy, yprime]
{{156, 128}, {80, 136}}
KEYbob = TFastPower[invKey, limit]
{{136, 128}, {80, 156}}
```